\documentclass{sig-alternate}

\usepackage{wrapfig}
\usepackage{graphicx}
\usepackage{epstopdf}
\usepackage{float}
\usepackage{color}
\usepackage{amsmath}
\usepackage{amssymb}
\usepackage{rotating}
\usepackage{url}
\usepackage{setspace}

\long\def\comment#1{}

\newtheorem{example}{Example}

\begin{document}

\conferenceinfo{KDD 2016 Workshop on Data Science for Food, Energy and Water,}{Aug 13-17, 2016, San Francisco}
	
	
\title{A Knowledge Ecosystem for the Food, Energy, and Water System}


\numberofauthors{3} 
%
\author{
	%
	%
	\alignauthor
	Praveen Rao\\
	\affaddr{Univ. of Missouri-Kansas City}\\
	\affaddr{Kansas City, MO}\\
	\email{raopr@umkc.edu}
	\alignauthor
	Anas Katib\\
	\affaddr{Univ. of Missouri-Kansas City}\\
	\affaddr{Kansas City, MO}\\
	\email{anaskatib@mail.umkc.edu}
	\alignauthor
	Daniel E. Lopez Barron \\
	\affaddr{Univ. of Missouri-Kansas City}\\
	\affaddr{Kansas City, MO}\\
	\email{dl544@mail.umkc.edu}
}

\maketitle


\newcommand{\ie}{{\em i.e.}}
\newcommand{\eg}{{\em e.g.}}
\newcommand{\etal}{{\em et al.\mbox{$\:$}}}
\newcommand{\noexists}{\mbox{$\backslash\!\!\!\!\!\;\exists$}}
\newcommand{\BigO}[1]{\mbox{${\cal O}(#1)$}}
\newcommand{\BigOmega}[1]{\mbox{$\Omega(#1)$}}
\newcommand{\BigTheta}[1]{\mbox{$\Theta(#1)$}}
\newcommand{\horizbar}{\rule{\linewidth}{.5mm}}
\newcommand{\ceiling}[1]{\left\lceil #1 \right\rceil}
\newcommand{\floor}[1]{\left\lfloor #1 \right\rfloor}
\newcommand{\faM}{\lfloor \alpha M \rfloor}
\newcommand{\C}[2]{{#1 \choose #2}}
\newcommand{\cardinal}[1]{\mbox{$\,\mid\!\! #1 \!\!\mid\,$}}
\newcommand{\xor}{\oplus}
\newcommand{\abs}[1]{\mbox{$\left|#1\right|$}}

\newcommand{\lpnorm}[2]{\mbox{$\left\|#1\right\|_{#2}$}}
\newcommand{\ltwonorm}[1]{\mbox{$\left\|#1\right\|_{2}$}}
\newcommand{\nrom}[1]{\lVert#1\rVert}
\newcommand{\lland}{\,\wedge\,}
\newcommand{\llor}{\,\vee\,}
\newcommand{\xby}{\!\times\!}
\newcommand{\combi}[2]{\mbox{$\left(\!\begin{array}{c}{#1}\\{#2}\end{array}\!\right)$}}
\newcommand{\tilda}{\verb+~+}
\newcommand{\pair}[1]{\mbox{$\langle #1 \rangle$}}
\newcommand{\edge}[1]{\mbox{$\overline{#1}$}}

\def\select#1#2{\mbox{$\sigma_{#1}(#2)$}}

\newcommand{\openbox}{\leavevmode
  \hbox to.77778em{%
  \hfil\vrule
  \vbox to.675em{\hrule width.6em\vfil\hrule}%
  \vrule\hfil}}
\newcommand{\qedsymbol}{\openbox}
\newcommand{\proofbegin}{{\em Proof. \ }}               
\newcommand{\proofend}{\hfill\qedsymbol\bigskip}        

\begin{abstract}

Food, energy, and water (FEW) are key resources to sustain human life and economic growth. There is an increasing stress on these interconnected resources due to population growth, natural disasters, and human activities. New research is necessary to foster more efficient, more secure, and safer use of FEW resources in the U.S. and globally. In this position paper, we present the idea of a knowledge ecosystem for enabling the semantic data integration of heterogeneous datasets in the FEW system to promote knowledge discovery and superior decision making through semantic reasoning. Rich, diverse datasets published by U.S. federal agencies will be utilized. Our knowledge ecosystem will build on Semantic Web technologies and advances in statistical relational learning to (a) represent, integrate, and harmonize diverse data sources and (b) perform ontology-based reasoning to discover actionable insights from FEW datasets.
\end{abstract}

\section{Introduction}

Food is required for the survival of mankind and is highly interconnected with energy and water resources. Inadequate supply of any of these resources will hamper the sustainability of human life and economic development. There is an increasing stress on these resources due to population growth, natural disasters like droughts, and human activities such as urbanization and industrialization~\cite{GHI2012}. There is a pressing need to foster more efficient, more secure, and safer use of these resources in the U.S. and globally.

There is a plethora of information related to the FEW system in the datasets published by the United States Department of Agriculture (USDA), the National Oceanic and Atmospheric Administration (NOAA), the U.S. Geological Survey (USGS), and the National Drought Mitigation Center (NDMC). Unfortunately, these datasets are in formats that cannot be easily integrated (\eg, Microsoft Excel, CSV, XML, JSON). Moreover, there is a lack of rich ontologies to model the domain knowledge of the FEW system. As a result, these datasets cannot be consumed by intelligent software systems for knowledge discovery and decision making. Today, Semantic Web technologies such as the Resource Description Framework~\cite{RDF}, the SPARQL query language~\cite{SPARQL1.1}, and the Web Ontology Language (OWL)~\cite{OWL2} are becoming key drivers for success in domain-specific applications that require data integration and knowledge management as well as publishing knowledge on the Web~\cite{LinkedData09}. These technologies are enabling semantic reasoning in domains such as biopharmaceuticals, defense and intelligence, and healthcare. Companies have adopted them for different use cases such as data aggregation (\eg, Pfizer~\cite{Pfizer}), publishing open data on the Web and providing better quality search results (\eg, Newsweek, BBC, The New York Times, Best Buy)~\cite{MITERMIT}. More recently, companies such as Microsoft, Google, and Yahoo! are employing knowledge graphs to provide higher quality search results and recommendations to users. Today, some of the popular knowledge graphs (\eg, YAGO2~\cite{Hoffart2011}, Wikidata~\cite{Wikidata2014}) are represented in RDF.


We posit that advances in Semantic Web technologies and statistical relational learning provide a timely and unique opportunity to advance the state-of-the-art in the FEW system. {\em In this position paper, we present the idea of a knowledge ecosystem for the FEW system to foster more efficient, more secure, and safer use of FEW resources.}  This knowledge ecosystem will enable (a) the semantic data integration of heterogeneous datasets in the FEW system; (b) the creation of an evolving knowledge base (KB) with rich ontologies; and (c) the ability to reason over the KB to discover actionable insights and validate scientific hypotheses.

\section{Background and Related Work}
\label{sec-background}

\subsection{RDF, SPARQL and OWL}
RDF is a standard model for representing data on the Web~\cite{RDF}. RDF uses International Resource Identifiers (IRIs) to name entities and their relationships and enables easy merging of different data sources. In RDF, a fact or assertion is represented as a (subject, predicate, object) triple. A set of RDF triples can be modeled as a directed, labeled graph. A triple's subject and object denote the source and sink vertices, respectively, and the predicate is the label of the edge from the source to the sink. An RDF statement denoted by (subject, predicate, object, context) is a quadruple. The context is used to capture the provenance or other relevant information of a triple. When the context is an IRI (a.k.a graph name), triples with the same context belong to the same RDF named graph. Below is an example of an RDF statement as a quadruple:

\noindent \texttt{<http://www.w3.org/People/Berners-Lee>} \texttt{foaf:givenname} ``\texttt{Timothy}'' \texttt{<http://www.w3.org/People/Berners-Lee/card>}.

SPARQL~\cite{SPARQL1.1} is the popular query language for RDF. It enables users to pose expressive graph queries on RDF data. One of the fundamental operations in RDF query processing is {\em Basic Graph Pattern} matching~\cite{SPARQL1.1}. Basic graph patterns (BGPs) are the building blocks in a SPARQL query and complex graph patterns can be expressed by combining BGPs through keywords such as OPTIONAL, FILTER, and UNION~\cite{SPARQL1.1}. Note that the triple patterns in the query contain certain nodes that are variables, which are prefixed by {\tt ?}. These variables are bound to actual RDF terms in the data (\eg, a resource with IRI, literal, blank node) during query processing via subgraph matching. Common variables within a BGP or across BGPs denote a join operation on the variable bindings of triple patterns.  An example of a SPARQL query is shown below, which uses two Friend of a Friend (FOAF) files to {\em find all those who know both Tim Berners-Lee and Ivan Herman and output them sorted by their nick names.}


\begin{footnotesize}
\begin{verbatim}
PREFIX foaf: <http://xmlns.com/foaf/0.1/>
SELECT ?c ?d ?e
FROM NAMED 
<http://www.w3.org/People/Berners-Lee/card>
FROM NAMED <http://www.ivan-herman.net/foaf.rdf>
WHERE {
GRAPH <http://www.w3.org/People/Berners-Lee/card> {
  ?a foaf:knows ?c .
  ?a foaf:name "Timothy Berners-Lee" .
  OPTIONAL { ?c foaf:depiction ?d .}}
GRAPH <http://www.ivan-herman.net/foaf.rdf> {
  ?b foaf:knows ?c .
  ?b foaf:name "Ivan Herman" .
  ?c foaf:nick ?e . }} 
ORDER BY ?e
\end{verbatim}
\end{footnotesize}

OWL 2~\cite{OWL2} is a knowledge representation language for defining ontologies. Using OWL, individuals, classes, properties, and relationships among them can be modeled for a domain. OWL reasoners~\cite{OWLReasoners} (\eg, Pellet, FaCT++, HermiT) can automate the discovery of information that humans would have missed. Given axioms and assertions, these reasoners can compute consequences~\cite{OWL2}--a useful feature for hypotheses verification and knowledge discovery. For example, using OWL 2 assertions, we can state that Timothy Berners-Lee and Ivan Herman belong to the class \textit{Person}. We can also state that different IRIs referring to Timothy Berners-Lee indicate the same person.

\subsection{Knowledge Management on the Web}

Several efforts have been made to publish data/metadata on the Web. For example, Data.gov~\cite{DataGov} provides a catalog of government and public datasets with rich metadata in RDF. Many of the actual datasets, however, are in their original formats provided by the publishers (\eg, Excel, CSV, XML, JSON). One noteworthy effort to map these datasets into machine-interpretable form is the TWC Data-gov Corpus~\cite{Ding2010}. However, the datasets necessary for the FEW knowledge ecosystem such as those published by the USDA, NOAA, USGS, and NDMC are not available in this corpus.

The Bioenergy Knowledge Discovery Framework~\cite{BioenergyKDF} is a DOE-funded effort for sustainable bioenergy industry by providing users with datasets, publications, and tools for fostering research and decision making in bioenergy. This framework allows users to analyze and visualize vast amounts of geospatial data for decision making. The datasets, however, are in formats that cannot be easily integrated with other diverse data sources. Also, the framework cannot conduct semantic reasoning. Recently, the International Food Policy Research Institute developed a tool called the Global Hunger Index~\cite{GHItool} to measure and track the hunger globally. Interestingly, the data are published as Linked Open Data in RDF and can be queried using SPARQL. However, the datasets are limited to hunger and nourishment.

\subsection{Markov Logic Network}

A Markov Logic Network (MLN)~\cite{Richardson2006} is regarded as one of the most flexible representations in statistical relational learning because it combines first-order logic and probabilistic graphical models. MLNs are widely used in natural language processing, entity resolution, hypertext classification, and information extraction and retrieval. Formally, a MLN is a KB defined by a set of pairs ($F$,$w$), where $F$ is a first-order formula that denotes a constraint and $w$ is a real-valued weight of the formula. Higher the weight, more likely is the constraint believed to be satisfied in the set of possible worlds. A formula with infinite weight is a hard constraint. Formulas can contradict. A world that violates a formula is less probable but not impossible. However, a world that violates a hard constraint has zero probability. Once the formulas and weights are learned~\cite{Singla2005}, probabilistic inference can be performed on the MLN by posing maximum a posteriori (MAP) and marginal inference queries. Efficient inferencing techniques have been developed (\eg, lifted inference~\cite{Milch2008,JhaGMS10}) including those that can operate on large KBs with millions of entities and facts by leveraging the scalability and efficiency of database systems (\eg, Tuffy~\cite{Niu2011}, ProbKB~\cite{Chen2014}).
\begin{example}
Consider a KB about smokers and friends with 3 formulas~\cite{Richardson2006}: $\forall x~Smoker(x) \implies Cancer(x)$;  $\forall x\forall y~Friends(x,y) \implies (Smoker(x) \iff Smoker(y))$; and $\forall x~Smoker(x)$ with weights 3.5, 1.0, and -1.0, respectively The first formula is a stronger constraint than the others and is of higher importance in the set of possible worlds. The third formula with -ve weight implies that a person is more likely to be a non-smoker. Using probabilistic inference, one can perform marginal inference queries on an entity (or all entities) and MAP queries. For example, we can compute the marginal inference queries such as $\Pr(Friends(Alice,Bob))$ and $\Pr(Cancer(x))$. We can also compute a MAP query such as $\arg\max\limits_x\Pr(Cancer(x))$.
\end{example}
\section{Our Position}
\label{sec-design}

\begin{figure}[tbh]
	\begin{center}
		\begin{tabular}{cc}
			\includegraphics*[width=3.3in, angle=0]{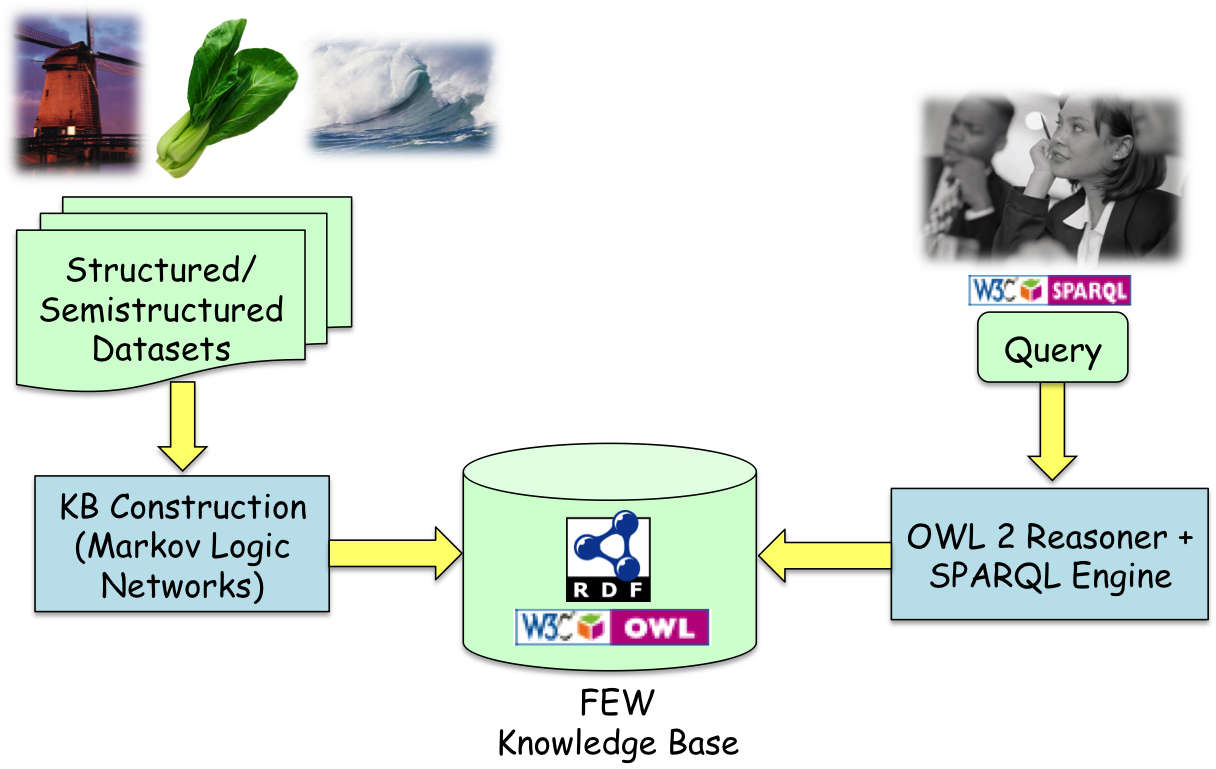}	
		\end{tabular}
	\end{center}
	\vspace*{-3ex}
	\caption{Proposed architecture}
	\vspace*{-2ex}
	\label{fig-FEW-architecture}
\end{figure}

\begin{figure}[!t]
	\begin{center}
		\begin{tabular}{c}
			\includegraphics*[width=2.8in, angle=0]{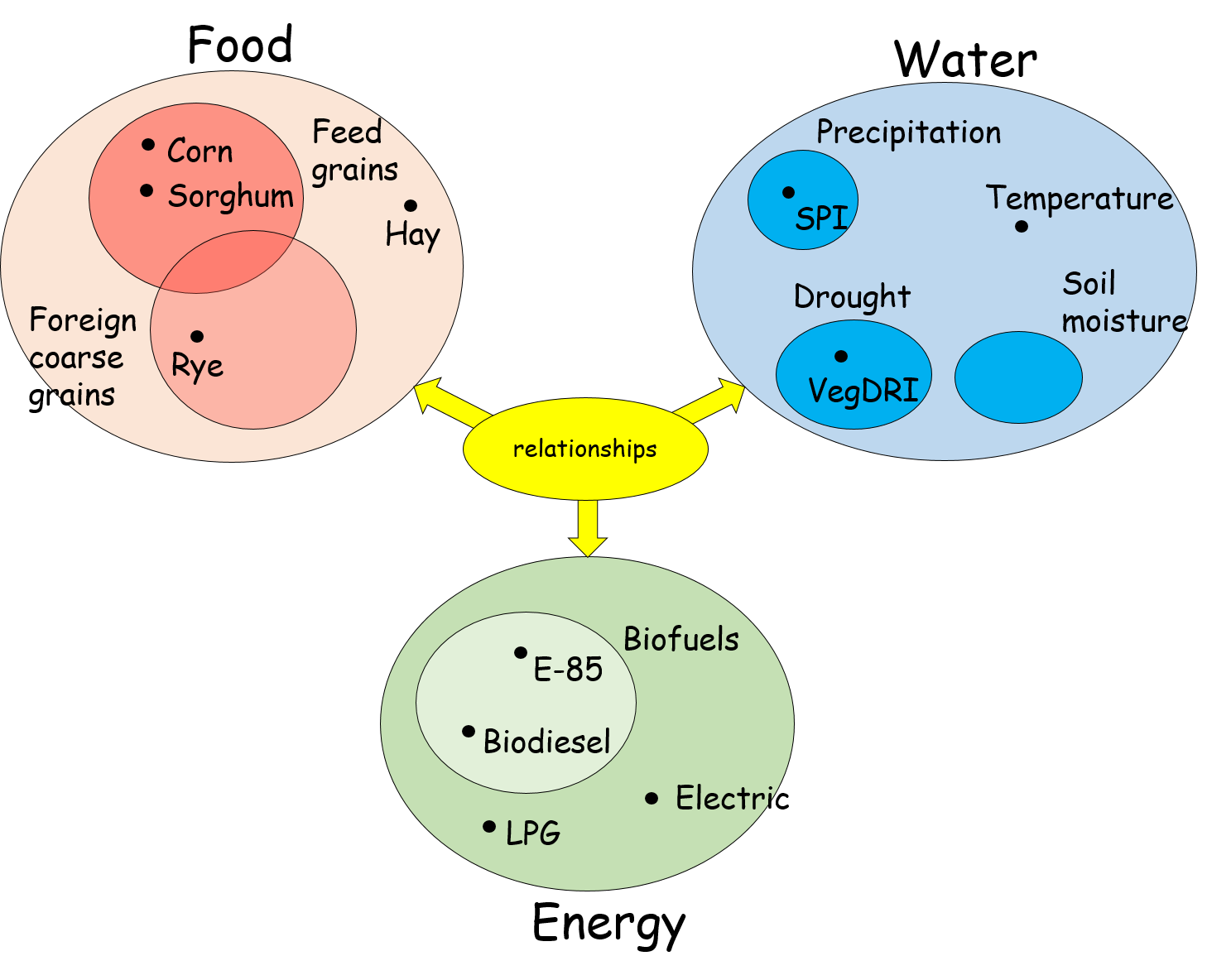} \\
			(a) Example of entities and concepts to model \\
			\includegraphics*[width=3.2in, angle=0]{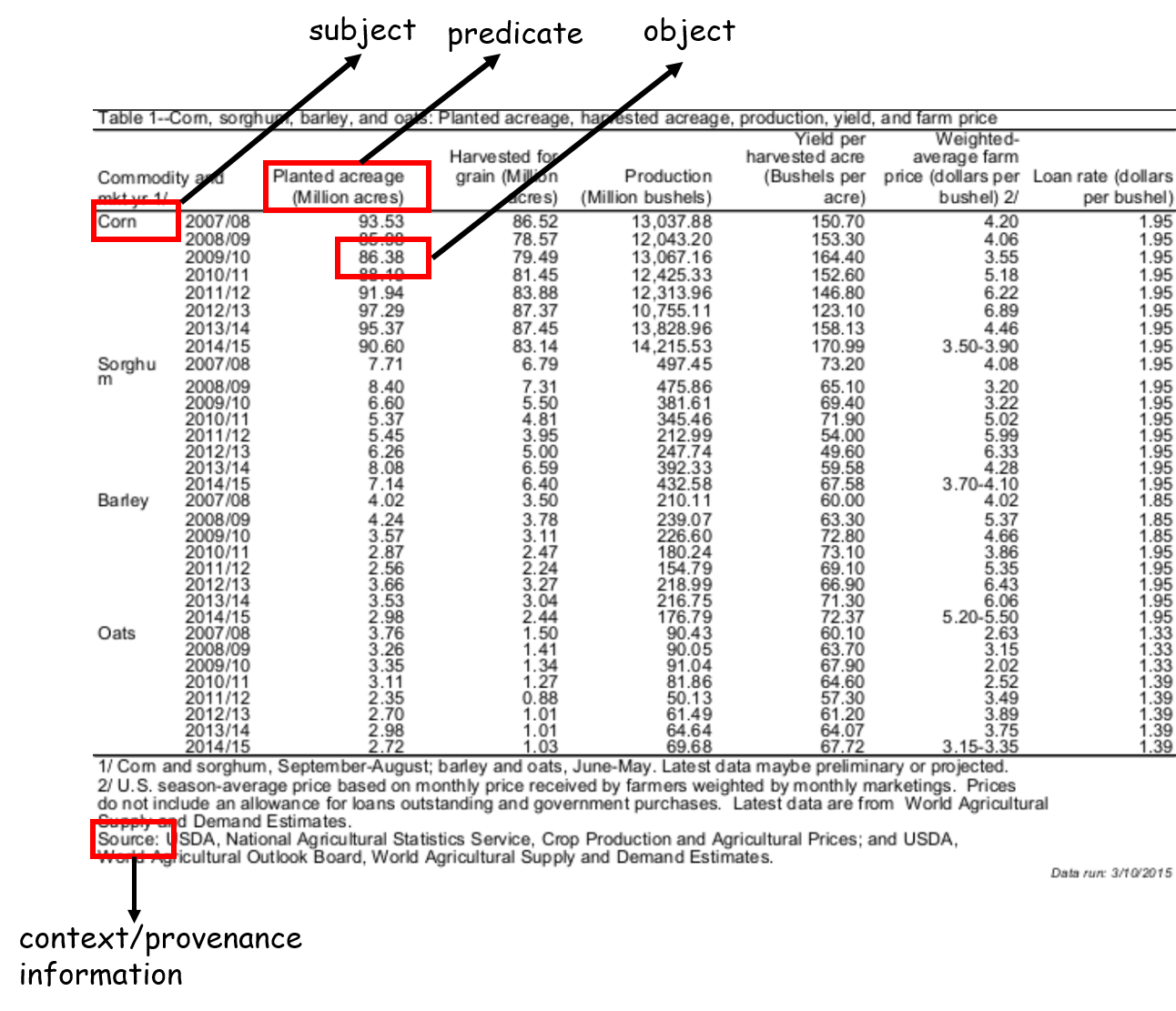} \\
			(b) Sample crop data (Source: USDA) \\
			\includegraphics*[width=3.0in, angle=0]{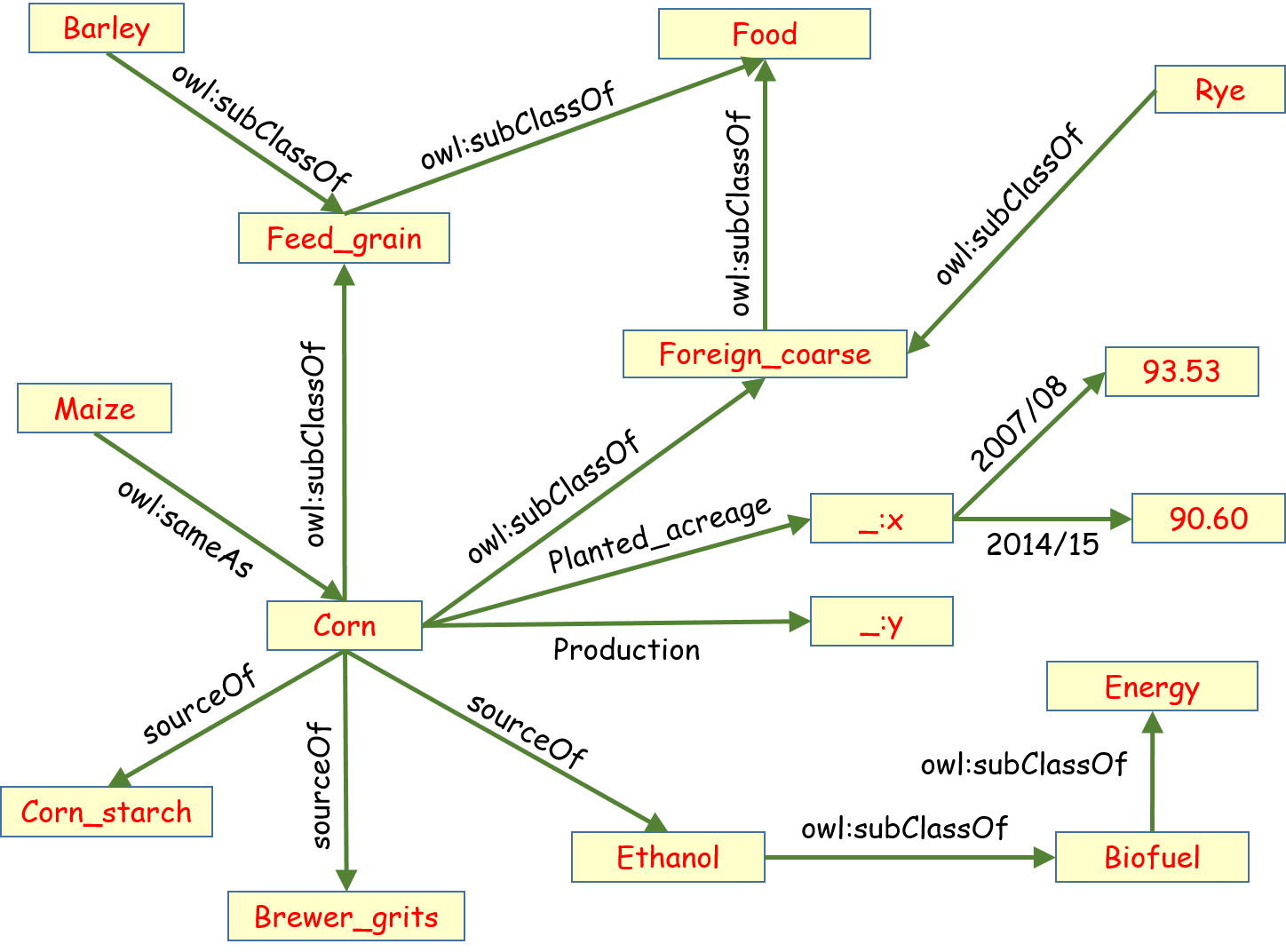} \\
			(c) RDF/OWL representation of FEW entities \& concepts
		\end{tabular}
	\end{center}
	\vspace*{-3ex}
	\caption{Examples on FEW data}
	\label{fig-examples}
	\vspace*{-2ex}
\end{figure}

\subsection{Proposed Architecture}
Given the lack of comprehensive KBs for the FEW system, we present the idea of a knowledge ecosystem for FEW, which can support knowledge management and discovery (through reasoning) using Semantic Web technologies and MLNs. This ecosystem will integrate a variety of data relevant to the FEW system such as agriculture productivity, vegetable/feed grains/livestock/dairy/meat consumption and production, food prices, land usage, irrigation, oil and gas production, biofuels, fertilizer usage and prices, income of farm households, drought conditions, soil moisture and precipitation, water quality, climate and weather, and others~\cite{USDAdata,NOAAdata,NDMC}. 

Our ultimate goal is to develop a system as shown in Figure~\ref{fig-FEW-architecture}. In this system, structured and semistructured datasets relevant to the FEW system will be processed using MLNs. Using techniques for entity resolution, information extraction/retrieval, and classification, we will infer rules in the FEW datasets and construct an integrated and harmonized KB. This KB will be represented using Semantic Web technologies, \ie, RDF and OWL 2 DL. An end-user (\eg, a data scientist) can perform semantic reasoning and query processing to (a) discover new insights from the FEW datasets and (b) validate scientific hypotheses.

\subsection{Construction of the FEW KB} 

Semantic data integration and knowledge management of the FEW datasets poses interesting technical challenges. Below is a motivating example that exemplifies the richness of entities and concepts in the FEW ecosystem.

\begin{example}
Consider corn, hay, and barley. Corn is a feed grain and an energy feed. It is the source for brewer grits, corn meal, corn starch, corn syrup, dextrose, ethanol, and others.  Hay is of two types: alfalfa and non-alfalfa. Barley is related to concepts such as barley feed, barley grain, and barley malting. Oil can be produced from different sources such as soybean, peanut, cottonseed, sunflower seed, linseed, flaxseed, and canola. Soybean and animal fat are used to produce biodiesel, which is a form of bioenergy. The Vegetation Drought Index (VegDRI) is affected by vegetation conditions, climate, land cover type, soil characteristics, and so on.
\end{example}

In Figure~\ref{fig-examples}(a), we show an example of a set of possible entities, classes, and relationships in the FEW system. To represent the learned KB with entities, relationships, rules, and facts, we will leverage popular vocabularies on the Web (\eg, Dublin Core~\cite{DCMI}, SKOS~\cite{SKOS}, Wikidata~\cite{Wikidata2014}) to define the RDF terms. We will consider a subset of concepts and entities in the FEW system to define the ontologies. Each RDF statement will be denoted by a (subject, predicate, object, context) quadruple. The context in a quadruple will capture the provenance information and serve as the name of the RDF graph. In Figure~\ref{fig-examples}(b), we show one possible way to map a (structured) crop data from USDA into a quadruple. 

We will leverage recent advances in entity resolution, classification, and information extraction via MLNs to infer properties in the OWL 2 vocabulary (\eg, \url{owl:sameAs}, \url{owl:equivalentProperty}, \url{owl:differentFrom}, \url{owl:subClassOf}) on entities (\ie, individual, class, or property) in the data. We will identify entities that should be modeled using the same IRIs. We will use scalable construction and reasoning techniques over probabilistic KBs~\cite{Niu2011,Chen2014}) to cope with very large FEW datasets leading to large number of RDF statements and OWL assertions. To represent the probabilities of the extracted rules/facts, we can use the concept of reification in RDF. Figure~\ref{fig-examples}(c) shows how the data and concepts could be modeled as an RDF named graph.

\subsection{Efficient Semantic Reasoning}

Several OWL reasoners have been developed over the years, which differ in the level of expressiveness of the ontologies and therefore, affect the performance of reasoning queries~\cite{OWLReasoners}. For example, OWL 2 DL is more expressive than OWL 2 EL. On the FEW KB, an end-user can use an OWL reasoner to automatically derive consequences and actionable insights from the KB, which a human could have missed. Here are two motivating examples for reasoning on the FEW KB.

\begin{example}
Suppose we have the following assertions in the FEW KB: ``Meat price increases when the demand increases. Lower crop production increases biofuel prices. There was drought in Jackson county, Missouri, in 20XX.'' Using an OWL reasoner, we can draw the conclusion that the drought in Jackson county, Missouri, lead to an increase in meat price and biofuel price in Kansas City in 20XX.
\end{example}

\begin{example}
This example is inspired by a quote from the NDMC's website~\cite{NDMCEval}: A rancher claims that they have received less than an inch of rainfall in the last few months. A researcher looks at the VegDRI value for the rancher's county and observes that it is ``green'' indicating good moisture levels. At first, the two statements appear to be contradictory. Using an OWL reasoner, the
researcher can verify that the two statements are consistent with each other, because the ranch had warm season grasses, which grew well even with little water.
\end{example}

Given the large number of FEW datasets available today, we anticipate the FEW KB to grow to billions of RDF statements and OWL assertions. One challenge is the performance of semantic reasoning via OWL 2 DL on such a large KB. We can use an OWL 2 DL reasoner such as Pellet~\cite{OWLReasoners}. We will map a reasoning query into complex, graph pattern queries on RDF quadruples in the FEW KB. This calls for efficient SPARQL query processing on RDF quadruples. For this purpose, we will employ our recently developed system called RIQ for efficient SPARQL query processing on RDF named graphs~\cite{WebDB2014Rao,RIQJWS2015}. RIQ will be used as the underlying RDF storage and query processing engine inside an OWL 2 DL reasoner. New query optimization techniques should be explored in RIQ to enable scalable OWL 2 DL reasoning on the FEW KB. Another avenue for research is to explore parallel RDF query processing using RIQ on large-scale data processing systems such as Apache Spark~\cite{Zaharia2012}.

\section{Concluding Remarks}
\label{sec-conclusion}

In this position paper, we argued for a knowledge ecosystem for the FEW system to foster more efficient, more secure, and safer use of FEW resources. This knowledge ecosystem will leverage Semantic Web technologies and advances in statistical relational learning to construct a KB over FEW datasets leading to semantic data integration and harmonization of heterogeneous datasets. Using ontology-based reasoning, our system can enable an end-user to reason over the FEW KB to discover actionable insights and validate scientific hypotheses in an automated manner. Our ultimate goal is to use the FEW knowledge ecosystem to facilitate superior decision making by FEW decision makers and stakeholders. One risk factor in our approach is the accuracy of statistical inference, which could lead to incorrect associations between entities in the KB.

\subsubsection*{Acknowledgments}
This work was supported in part by the National Science Foundation under Grant No. 1115871 and Grant No. 1620023.

\begin{scriptsize}
\bibliographystyle{abbrv}
\bibliography{bay}
\end{scriptsize}

\end{document}